# Disorder and critical phenomena

**B R Gadjiev**

International University for Nature, Society and Man, 19 Universitetskaya Street, 141980 Dubna, Russia

E-mail: gadjiev@uni-dubna.ru

**Abstract.** In the present paper we are discussing the influence of the fractal distribution of defects on the critical behavior of the system. We consider a case when the equation of motion for the order parameter contains influences of defects of random field and random temperature types at the same time and we show that it can lead to a change of distribution function from Gibbs distribution to Tsallis distribution or more general statistics of $q$-type. We are able to extend the Landau-Khalatnikov equation for the order parameter and represent it in the form of a differential equation of the fractional order. We deduce and we solve the corresponding renormalization group equation with the nonlinear dispersion law and we calculate the critical indices of the system.

PACs: 74.62.Dh, 75.10.-b, 77.80.Bh

## 1. Introduction

The theoretical description of spontaneous magnetic symmetry breaking in ideal crystals in the terms of the inverse Landau problem presupposes that the symmetry groups of paramagnetic $H_0$ and magnetic $H_1$ phases are known [1,2]. The group $H_1$ is a subgroup of the $H_0$, i.e. $H_1 \subset H_0$. If the groups $H_0$ and $H_1$ are known, it is possible to determine the finite matrix group $\Xi$, which describes transformation properties of the major order parameter $\eta$. In this case, in accordance with the symmetry arguments, the expansion of the free energy functional of the crystal contains an integer basis of the matrix group $\Xi$. Besides, the major order parameter is a certain linear combination of the coordinates of the system elements. Thus, it should be expected that the equation of motion for the order parameter would be rather complex in the general case. Even if small deviations of the order parameter from its equilibrium value are taken into account, the equation of motion for the order parameter can contain time derivatives of an arbitrarily high order [3, 4].

Moreover, there are always defects and impurities in real crystals. That is why the issue of the weak disorder influence on critical behaviour of the system appears quite natural.

Random distribution of defects in the structure may completely destroy large-scale fluctuations. As a result, either the singularities of thermodynamic functions disappear or a shift of the critical temperature takes place.

However, the conditions of the crystal growth may lead to inhomogeneous distribution of defects. In this case, a new critical regime may arise due to impurities that are characterized by new universal critical indices.

The critical behaviour of the system near the critical point is determined by large-scale fluctuations. The correlation length is the only scale parameter that exists in the system near the phase transition point. The correlation length increases approaching the critical point, and becomes infinite in the phase transition point. First, the fluctuation correlation length becomes much longer than the lattice characteristic distance. Further approaching the critical point, the correlation length becomes longer than the mean distance between impurities. Thus, the effective concentration of impurities measured in the scales of the correlation length, becomes large. A situation occurs at certain conditions when a new critical regime arises sufficiently near the critical point. This new regime is described by new universal critical indices. The Harris criterion allows us to predict in what conditions the impurities are significant to the critical behaviour and in what conditions are not. If the critical index of the ideal crystal specific heat is positive, the weak disorder influences the critical behaviour. Otherwise, the weak disorder has no influence on the critical behaviour. The analysis of the weak disorder effect on the system critical behaviour in the framework of the renormalization group method and $\varepsilon$ - expansion is given in [5]. It was shown that that the critical indices of the system do not change (i.e. the universality class remains unchanged) [5]. In [5] by using the broken replica symmetry method it was shown that in the case of weak quenched disorder the universality class of the system may change.

We have studied in the paper the effect of the weak disorder on the critical behaviour of the system in the case of fractal distribution of defects. We present a fractional generalization of the dynamic equation for the major order parameter and obtain the nonlinear dispersion law. We show that in this case a new critical regime appears.

**2. The influence of the defects of the random field and random temperature types and the Fokker-Plank equation for the distribution function of the order parameter**

In the continuous description of the system in the terms of the Ginzburg-Landau Hamiltonian the presence of impurities may reveal in weak random space fluctuations of the given temperature (i.e. the presence of impurities leads to space fluctuations of the effective local temperature of the transition) or in weak space distortions of the crystal local symmetry (i.e. the presence of impurities leads to space fluctuations inside the crystal field). Let us introduce the notation $f(\eta) = -\dfrac{\delta F[\eta]}{\delta \eta}$. Then, the equation of motion for the order parameter $\eta$, with the fluctuation of the local temperature and local field being present, may be given in the form

$$\frac{\partial \eta}{\partial t} = f(\eta) + u(\eta)\xi(t) + \varsigma(t), \qquad (1)$$

where $\eta(t)$ is a stochastic variable, $u(\eta)$ is arbitrary function and $\xi(t)$ and $\eta(t)$ are uncorrelated noises with mean value equal to zero

$$\langle \xi(t)\xi(t') \rangle = 2M\delta(t-t'), \qquad (2)$$

$$\langle \varsigma(t)\varsigma(t') \rangle = 2A\delta(t-t'), \qquad (3)$$

where $M > 0$ and $A > 0$ are amplitudes of the multiplicative and additive noises, respectively.

The Fokker-Plank equation for probability density $p(\eta,t)$, connected to equation (1) in the Stratonovich calculus has the form [6]

$$\frac{\partial p(x,t)}{\partial t} = -\frac{\partial (f(\eta)p(\eta,t))}{\partial \eta} + M\frac{\partial}{\partial \eta}\left(u(\eta)\frac{\partial}{\partial \eta}(u(\eta)p(\eta,t))\right) + A\frac{\partial^2 p(\eta,t)}{\partial \eta^2} \qquad (4)$$

Suppose that

$$f(\eta) = -\tau u(\eta)\frac{\partial u(\eta)}{\partial \eta}.$$

In this case it is easy to show that the stationary solution of this equation takes the form [6]

$$p_{st}(x) = p_0 A^{-\frac{\tau+M}{2M}} \left(1 + \frac{M}{A} u^2(\eta)\right)^{-\frac{\tau+M}{2M}}, \qquad (5)$$

and, consequently,

$$p_{st}(x) = \frac{1}{Z}\left(1-(1-q)u^2(\eta)\right)^{\frac{1}{1-q}} \qquad (6)$$

where $q = \dfrac{\tau + 3M}{\tau + M}$ and $\dfrac{M}{A} = q - 1$.

It is clear that $u^2(\eta) = -\dfrac{2}{\tau}F[\eta]$. Thus, the multiplicative noise effect leads to the nonextensivity system [6, 7].

We have to mark that as inherently $q = \dfrac{3M + \tau}{M + \tau}$, at $M = 0$ we obtain $q = 1$. In this case the stationary distribution $p_{st}(\eta)$ is the Boltzmann-Gibbs distribution and, consequently, it is possible to state that the system nonextensivity is connected to the effect of multiplicative noises on the system. Besides, the obtained distribution at large $u^2(\eta)$ results in the power law distribution. If $q \neq 1$ and $u^2(\eta)$ is bounded, the behaviour occurs that is intermediate between the exponential and power law one. It should be stressed that physically the presence of the multiplicative noise means that some positions in the process of the crystal growth are more preferable for the location of impurities. The saturation effect of chemical bonds leads to the fact that the number of bonds connected with the given node is always finite. Consequently, the distribution of the order parameter is neither the Boltzmann-Gibbs distribution nor the power law distribution, but the distribution of the $q$–type (5).

### 3. Presumable generalizations of the equation of motion for the order parameter

For this purpose, let us introduce the non local function $K(\tau)$ that characterizes time dispersion connected to the presence of weak disorder in the structure [4]. The latter means that, independent of time, the solution of the dynamic equation of motion for the major order parameter coincides with that of the statistical problem for ideal crystal. The generalized equation of motion for the major order parameter in the structure with weak disorder is represented in the form

$$\gamma \eta(x,t) = -\int_0^t K(t-\tau)\frac{\delta H(\eta)}{\delta \eta}d\tau \qquad (7)$$

Thus, the introduction of the $K(\tau)$ function into equation (7) characterizes the influence of the disorder on the generalized force $-\dfrac{\delta H(\eta)}{\delta \eta}$, whose action produces motion of the order parameter $\eta$. Physically, the $K(\tau)$ function describes the effect of heterogeneities of the crystal potential energy induced by the defects in the structure on the relaxation process of the order parameter to the equilibrium state.

If the paramagnetic phase symmetry group contains inversion $H = \dfrac{\alpha}{2}\eta^2 + \dfrac{\beta}{4}\eta^4$, where $\alpha$ is the reduced temperature and $\beta > 0$.

If $K(\tau) = const$, it may be shown by simple derivation that equation (7) takes the form Landau-Khalatnikov equation

$$\gamma_0 \frac{\partial \eta(x,t)}{\partial t} = -\frac{\delta H(\eta)}{\delta \eta} \qquad (8)$$

In systems without memory $K(t-\tau) = g_0 \delta(t-\tau)$ equation (7) takes the form of the nonlinear algebraic equation

$$\gamma \eta(x,t) = -g_0 \frac{\delta H(\eta)}{\delta \eta}. \tag{9}$$

or

$$\gamma \eta(x,t) = -g_0 \left[\alpha \eta + \beta \eta^3\right]$$

or

$$\left(\alpha + \frac{\gamma}{g_0}\right) \eta(x,t) + g_0 \beta \eta^3 = 0. \tag{10}$$

Hence, in this case the symmetry of the low-temperature phase remains changeless; however, a shift of the critical temperature occurs in the system and the state of the system depends on time.

Let us assume that the presence of defects of the random field and random temperature types in the structure leads to

$$K(t-\tau) = K_0 \left[1 - (1-q)\frac{(t-\tau)}{\tau_0}\right]^{\frac{1}{1-q}} \tag{11}$$

with $q \geq 1$. Then equation (7) takes the form

$$\gamma \eta(x,t) = -K_0 \int_0^t \left[1 - (1-q)\frac{(t-\tau)}{\tau_0}\right]^{\frac{1}{1-q}} \frac{\delta H[\eta(x,\tau)]}{\delta \eta} d\tau \tag{12}$$

At $q \neq 1$ and big enough $(t-\tau)$ equation (7) is written in the form

$$\gamma' \eta(x,t) = -\int_0^t \frac{1}{(t-\tau)^{\frac{1}{q-1}}} \frac{\delta H[\eta(x,\tau)]}{\delta \eta} d\tau, \tag{13}$$

where $\gamma' = \frac{\gamma}{K_0}(1-q)^{\frac{1}{q-1}}$. We introduce the notation $\sigma = \frac{1}{q-1}$, and let $0 < \sigma < 1$, hence $q > 2$.

Using the definition of the Riemann-Liouville fractional derivative [8]

$$\left(D_{a+}^\sigma f\right)(x) = \frac{1}{\Gamma(1-\sigma)} \frac{d}{dx} \int_a^x \frac{f(t)}{(x-t)^\sigma},$$

equation (13) may be reproduced in the form of the differential equation of the fractional order

$$\gamma_0' \frac{\partial^{1-\sigma}}{\partial t^{1-\sigma}} \eta(x,t) = -\frac{\delta H[\eta(x,t)]}{\delta \eta}, \tag{14}$$

where $\gamma_0' = \frac{\gamma'}{\Gamma(1-\sigma)}$ and $\Gamma(x)$ — is the Euler gamma function.

Let us consider the limit $q \to 1$. In this case

$$\lim_{q \to 1} K(t-\tau) = \lim_{q \to 1} K_0 \left[1 - (1-q)\frac{(t-\tau)}{\tau_0}\right]^{\frac{1}{1-q}} = K_0 e^{-\frac{(t-\tau)}{\tau_0}} \tag{15}$$

Then equation (7) takes the form

$$\gamma \eta(x,t) = -\int_0^t e^{-\frac{(t-\tau)}{\tau_0}} \frac{\delta H(\eta)}{\delta \eta} d\tau \tag{16}$$

Using Taylor's expansion of the exponential function, we may write

$$\int_0^t e^{-\frac{(t-\tau)}{\tau_0}} \frac{\delta F(\eta)}{\delta \eta} d\tau = \left[ D^{-1} + \frac{1}{\tau_0} D^{-2} + \frac{1}{\tau_0^2} D^{-2} + ... \right] \frac{\delta H(\eta)}{\delta \eta} = \frac{\tau_0}{1+\tau_0 D} \frac{\delta H(\eta)}{\delta \eta} \quad (17)$$

By using equation (17) equation (16) takes the form

$$\gamma \left(1 + \tau_0 \frac{\partial}{\partial t}\right) \eta(x,t) = -\tau_0 \frac{\delta H(\eta)}{\delta \eta} \quad (18)$$

In this case the renormalization of the effective Hamiltonian coefficients and a shift of the critical temperature occur. Nevertheless, the presence of defects in the structure does not change the symmetry of the high-symmetry and commensurate phases.

Thus, if the distribution of defects in the structure is fractal, the equation of motion for the order parameter is described by the fractional differential equation.

Taking into account spatial heterogeneities of the order parameter we may show that the fractional differential equation for the order parameter leads to the nonlinear dispersion law [9]. Let us consider the free energy functional of the system with weak disorder, accounting for non local space heterogeneities. In this case it is suitable to represent the equation of motion for the order parameter in the form

$$\int_0^t G(t-t') \frac{\partial \eta(r,t')}{\partial t'} dt' = -\int_0^r dr' V(r-r') \frac{\partial^2 \eta(r-r')}{\partial r'^2} dr' - \int_0^r dr' U(r-r')(\alpha \eta(r',t) + \beta \eta^3(r',t)) dr' \quad (19)$$

Let us assume that $U(r-r') = \delta(r-r')$. Then we obtain from equation (19)

$$\int_0^t G(t-t') \frac{\partial \eta(r,t')}{\partial t'} dt' = -\int_0^r dr' V(r-r') \frac{\partial^2 \eta(r-r')}{\partial r'^2} dr' - \alpha \eta(r',t) - \beta \eta^3(r',t) \quad (20)$$

Let

$$G(t-t') = \frac{1}{\Gamma(1-\nu)} \frac{1}{(t-t')^\nu} \quad (21)$$

and

$$V(r-r') = \frac{1}{\cos\left(\frac{\pi \varepsilon}{2}\right) \Gamma(2-\varepsilon)} \frac{1}{|r-r'|^{\varepsilon-1}} \quad (22)$$

where $0 < \nu < 1$ and $1 < \varepsilon < 2$.

After substituting (21) and (22) into equation (20) we obtain

$$\frac{\partial^\nu \eta(r,t)}{\partial t^\nu} = -\kappa \frac{\partial^\varepsilon \eta(r,t)}{\partial r^\varepsilon} - \alpha \eta(r,t) - \beta \eta^3(r,t), \quad (23)$$

where for $\frac{\partial^\varepsilon}{\partial r^\varepsilon}$ we used the Caputo definition of the fractional derivative. Using the Fourier transformation we obtain the nonlinear dispersion law

$$(i\omega)^\nu = (k^2)^{\frac{\varepsilon}{2}} + \alpha.$$

## 4. Renormalization group analysis

All the renormalization group calculations start with a classical effective Hamiltonian $\overline{H}$, which is a functional of the «spins» $\sigma_{\vec{k}}$. Notice that the $\overline{H}$ depends on the integer basis of invariants of the matrix group $\Xi$.

In case of the fractal distribution of defects in the structure the partition function has the form [10]

$$Z = \int \left( \prod_{\vec{k}} d^n \sigma_{\vec{k}} \right) \exp_q \left( \mathrm{H}[\sigma_{\vec{k}}] \right) \qquad (24)$$

where

$$\exp_q(x) = [1 + (1-q)\overline{\mathrm{H}}]^{\frac{1}{1-q}}$$

We note that

$$\lim_{q \to 1} \exp_q(x) = \lim_{q \to 1} [1 + (1-q)\overline{\mathrm{H}}]^{\frac{1}{1-q}} = e^{\overline{\mathrm{H}}}.$$

If the symmetry group of the paramagnetic phase contains symmetry element of space inversion, the effective Hamiltonian of the system is given in the expression

$$\overline{\mathrm{H}} = -\frac{1}{2} \int_{\vec{k}} r(k) \sigma_{\vec{k}} \sigma_{-\vec{k}} - \frac{u}{4} \int_{\vec{k}} \int_{\vec{k}'} \int_{\vec{k}''} \sigma_{\vec{k}'} \sigma_{\vec{k}''} \sigma_{\vec{k}} \sigma_{-\vec{k}-\vec{k}'-\vec{k}''} \qquad (25)$$

where $\int_{\vec{k}} = \frac{1}{\Omega_0} \int d^d k$.

The integral over $\vec{k}$ is over the appropriate Brillouin zone, which is of volume $\Omega$. For a simple cubic lattice of spacing $a$, $\Omega_0$ is equal to $\Omega_0 = \left(\frac{2\pi}{a}\right)^d$.

The basic idea of the renormalization group iteration procedure is to perform the integral in (24) in steps. To do it, we presuppose that the integration in the moment in the initial system is accomplished in the region $0 < |\vec{k}| < \Lambda$. The new system is obtained by the integration of high-frequency modes with $\frac{\Lambda}{b} < |\vec{k}| < \Lambda$, $b > 1$. Function $\sigma_{\vec{k}}$ is given in the form [10]

$$\sigma_{\vec{k}} = \sigma_{0,\vec{k}} + \sigma_{1,\vec{k}},$$

where

$$\sigma_{1,\vec{k}} = \begin{cases} \sigma_k, & \text{if } \frac{\Lambda}{b} < |\vec{k}| < \Lambda \\ 0, & \text{if otherwise} \end{cases},$$

$$\sigma_{0,\vec{k}} = \begin{cases} \sigma_k, & \text{if } 0 < |\vec{k}| < \frac{\Lambda}{b} \\ 0, & \text{if otherwise} \end{cases}.$$

Thus, we first integrate over the variables $\sigma_{\vec{k}}$ which satisfy $\frac{\Lambda}{b} < |\vec{k}| < \Lambda$, $b > 1$.

Denoting

$$\exp_q(\mathrm{H}_1) = \int \left( \prod_{\frac{\Lambda}{b} < |\vec{k}| < \Lambda} d^n \sigma_{\vec{k}} \right) \exp_q(\mathrm{H}[\sigma_{\vec{k}}]) \qquad (26)$$

we can therefore write

$$Z = \int \left( \prod_{|\vec{k}| < \frac{\Lambda}{b}} d^n \sigma_{\vec{k}} \right) \exp_q(\mathrm{H}_1), \qquad (27)$$

where now $H_1$ is a functional of all variables $\sigma_{\vec{k}}$ with $|\vec{k}| < \frac{\Lambda}{b}$. We have thus eliminated from the problem the fluctuations which had the shortest wave-lengths.

We now aim to set (27) to the form similar to (24). This is done in two steps: we first preserve the space density of degrees of freedom, by rescaling all space vectors by a factor $b$,

$$\vec{k} \Rightarrow \vec{k}' = b\vec{k} .\tag{28}$$

We then rescale all spin variables, in order to preserve the basic spin fluctuation magnitude,

$$\sigma_{\vec{k}} \Rightarrow \sigma'_{b\vec{k}} = \sigma'_{\vec{k}'} = \varsigma^{-1}\sigma_{\vec{k}} \tag{29}$$

Introducing these two rescaling into $H_1$ leads to a new effective Hamiltonian,

$$\overline{H}'[\sigma'_{\vec{k}'}] = \overline{H}_1[\sigma_{\vec{k}}] \tag{30}$$

In the applications to short range isotropic forces, the spin rescaling factors $\varsigma$ are chosen so that the coefficient of $\vec{k}^2$ in $r(k)$ remains equal to unity.

After the rescaling factors $\varsigma$ are chosen, we can compare $\overline{H}'$ with $\overline{H}$, and write down a series of recursion relations for the coefficients $r(k)$, $u$, etc. These form a representation of the renormalization group transformation

$$\overline{H}' = R\overline{H} .\tag{31}$$

The next step is to follow the "flow" of $\overline{H}$, in the multi-dimensional Hamiltonian space, under the transformation $R$. For certain values of the parameters in $\overline{H}$, this flow will go to a fixed point, $\overline{H}^*$, at which

$$\overline{H}^* = R\overline{H}^* .\tag{32}$$

In most cases of interest, we shall find then one more fixed point.

A direct way to obtain the critical exponents is to linearize (31) near (32),

$$R(\overline{H}^* + \mu Q) = \overline{H}^* + \mu L Q + O(\mu^2) \tag{33}$$

And to find the eigenvalues $\Lambda_i$ and the eigenoperators $Q_i$ of the linear operator $L$,

$$LQ_i = \Lambda_i Q_i = b^{\lambda_i} Q_i .\tag{34}$$

If $\lambda_i > 0$, the coefficient of $Q_i$ will become larger under iterations. The operator $Q_i$ is then called *relevant*. If $\lambda_i < 0$, this coefficient will decay, and $Q_i$ is called irrelevant. If $\lambda_i < 0$ , $Q_i$ is called marginal [9].

It is easy to show by the direct calculation that the substitution of the Boltzmann-Gibbs distribution to the Tsallis distribution does not change the critical behaviour of the system, if we proceed from the effective Hamiltonian of the ideal crystal.

Accounting for the fractal distribution of defects in the structure, let us consider the influence of the weak disorder on the phase transition from the paramagnetic phase to the magnetic one. In this case the phase transition is described by the Heisenberg Hamiltonian, taking into account the nonlinear dispersion law. Thus, the effective Hamiltonian, necessary for the renormalization group analysis, has the form:

$$H[\sigma(\vec{q})] = -\frac{1}{2}\int_{\vec{q}} \omega(\vec{q})\sigma(\vec{q})\sigma(-\vec{q}) - \int_{\vec{q}}\int_{\vec{q}'}\int_{\vec{q}''} \sigma(\vec{q})\sigma(\vec{q}')\sigma(\vec{q}'') \tag{35}$$

where $\omega(q) = \gamma + q^{1+\nu}$ is taken into account.

To obtain the renormalization group equation, we write the statistical sum in the form:

$$Z = \int \left(\prod_{\vec{q}} d^n \sigma(\vec{q})\right)\left(e_q^{H[\sigma(\vec{q})]}\right) . \tag{36}$$

Let us specify

$$e^{H_1} = \int \left( \prod_{\Lambda/q<|q|<\Lambda} d^n\sigma(\vec{q}) e_q^{H[\vec{\sigma}(\vec{q})]} \right) \tag{37}$$

The invariance condition of the statistical sum in relation to the renormalization group transformations demands that

$$Z = \int \left( \prod_{|q|<\Lambda/b} d^n\sigma(\vec{q}) \right) \left( e_q^{\bar{H}_1[\sigma(\vec{q})]} \right), \tag{38}$$

where functional $H_1$ depends on variables $\sigma(\vec{q})$ с $|\vec{q}| < \Lambda/b$.

Besides, the invariance condition needs scale transformation of momenta $\vec{q}$ and fields $\sigma(\vec{q})$:

$$\vec{q} \Rightarrow \vec{q}' = b\vec{q},$$
$$\sigma(\vec{q}) \Rightarrow \sigma(\vec{q}) = \sigma'(\vec{q}') = \sigma'(b\vec{q}) = \xi^{-1}\sigma(\vec{q}). \tag{39}$$

Assuming the presence of the fixed point, we obtain the renormalization group equations:

$$r' = \xi^2 b^{-d}[r + 4(n+2)uA(r) + ...],$$
$$u' = b^{-3d}\xi^4[u - 4(n+8)u^2 C(r) + ...]. \tag{40}$$

We have introduced the following values here:

$$A(r) = \int_{\vec{q}} \omega(\vec{q}), \tag{41}$$

$$C(r) = \int_{\vec{q}} \omega^{-2}(\vec{q}), \tag{42}$$

where integration in the momenta is produced in the region $\Lambda/b < |\vec{q}| < \Lambda$ in the $d$-dimension space. The parameter $\xi$ is derived from the condition of equality to unit of the coefficient at $\vec{q}$ in $\omega(q)$. It produces $\xi^2 b^{-d-\gamma} = 1$, from which it follows that $\xi^2 b^{-d} = b^\gamma$, where $\gamma = 1 + \nu$.

Consequently

$$\xi^4 b^{-3d} = b^{-2\gamma+d}.$$

Thus, the renormalization group equation has the form:

$$r' = b^\gamma [r + 4(n+2)uA(r) + ...],$$
$$u' = b^{-d+2\gamma}\xi^4[u - 4(n+8)u^2 C(r) + ...]. \tag{43}$$

Let us assume that $\gamma = \frac{3}{2} + \varepsilon$, where $\varepsilon \in (0, 1/2)$. Let us determine the critical indices in space with dimension $d = 3$. In this case, we obtain from equation (43)

$$r' = b^{(3/2+\varepsilon)}[r + 4(n+2)uA(r) + ...],$$
$$u' = b^{2\varepsilon}[u - 4(n+8)u^2 C(r) + ...]. \tag{44}$$

Direct calculation shows that $C(r) \approx K_d \ln b$, and in the fixed point

$$u^* = \frac{2\varepsilon}{4(n+8)K_d}. \tag{45}$$

Taking into account that $A(r) \approx -rC(0)$, we obtain

$$r' = b^{(3/2+\varepsilon)}[r - 4(n+2)ruC(0) + ...]. \tag{46}$$

The linearization of the latter equation in the fixed point $u^*$ gives

$$\Delta r' = b^{1/\nu} \Delta r,$$

that allows the determination of the critical index

$$\nu = \left( \frac{3}{2} + \varepsilon - \frac{2(n+2)\varepsilon}{n+8} \right)^{-1} \qquad (47)$$

where $n$ is a number of components of the order parameter.

According to the similarity relation, the critical index of the magnetic susceptibility is $\gamma = 2\nu$.

Thus, we can state that critical indices are the function of the system fractal dimension in the structures with fractal distribution of defects.

**5. Conclusion**
It should be noted that in the usual approach of the renormalization group and $\in$–expansion $\in = 4 - d$ is used as a small parameter, where $d$ is the space dimension. We have succeeded to use the remainder modulus between the fractal dimension of the structure and the nearest whole number as a small parameter.

Besides, in our approach the structure fractality is introduced by the function $K(x - x')$ which is characterized by the interaction of the defects network on the dynamics of the order parameter. The determination of the function $K(x - x')$ in various fractal dimensions leads to various values of the order of the fractional differential equation that describes the motion of the order parameter. It changes the relaxation law of the order parameter and leads to the nonlinear dispersion law. The renormalization group analysis shows that in this case a new critical regime appears, together with the dependence of critical values on the structure fractal dimension. In particular, our result allows us to understand the nature of changes in temperature dependences of thermodynamic functions with changes of defects concentration which is widely observed in mixed crystals with [11].